# Intensity-based dynamic speckle method using JPEG and JPEG2000 compression

Elena Stoykova,[1*] Blaga Blagoeva,[1] Natalya Berberova-Buhova[1,] Mikhail Levchenko,[1] Dimana Nazarova,[1] Lian Nedelchev,[1] Joongki Park[2]

[1]*Institute of Optical Materials and Technologies, Bulgarian Academy of Sciences, Acad. Georgi Bonchev Str., Bl.109, 1113 Sofia, Bulgaria*
[2]*Electronics and Telecommunications Research Institute, 218 Gajeong-ro, Yuseong-gu, Daejeon, 34129, Republic of Korea*
*\*Corresponding author: estoykova@iomt.bas.bg*

**Statistical processing of speckle data enables observation of speed of processes. In intensity-based pointwise dynamic speckle analysis, a map related to speed's spatial distribution is extracted from a sequence of speckle patterns formed on an object under coherent light. Monitoring of time evolution of a process needs storage, transfer and processing of a large number of images. We have proposed lossy compression of these images using JPEG and JPEG2000 formats. We have compared the maps computed from non-compressed and decompressed synthetic and experimental images, and we have proven that both compression formats can be applied in the dynamic speckle analysis. © 2022 Optica Publishing Group**

## 1. INTRODUCTION

Dynamic speckle method (DSM) is growing in popularity due to simple data acquisition, high sensitivity and applicability to three-dimensional (3D) objects [1-3]. The method is evolving rapidly thanks to advances in computers and 2D optical sensors [4]. Most frequently, the DSM relies on i) acquisition of varying in time speckle patterns formed on the surface of diffusely reflecting objects under laser illumination and ii) statistical processing of speckle intensity data. The aim is obtaining information about the speed of the process that has caused the change in the speckle patterns. To name a few of the reported applications, the DSM has been applied for monitoring of blood flow perfusion in human tissues [5-7], penetration of cosmetic ingredients in human skin [8], ear biometrics [9], bacterial response [10,11], plants growth and leaves chemical contamination [12,13], seeds viability [14,15], animal reproduction [16], food quality assessment [17-19], and drying of paints, coatings and polymer thin films [20,21] .

The strong impact of micro changes in topography or refractive index on speckle formation ensures high DSM sensitivity for evaluation of the speed of processes. The speed, which may vary on the object surface, is encoded in the rate of speckle intensity fluctuations in time. A 2D map is built for representing the speed distribution across the object by pointwise computing of a given statistical parameter from a sequence of correlated in time speckle images. The map is mostly known as an activity map since it presents an instant picture of areas of faster or slower intensity changes across the object surface. Accuracy of the DSM increases while the temporal resolution decreases with the number of images, $N$, needed for a single map. This number depends on the tested object, but usually it is rather large, e.g. it may reach 64, 128 or 256 images. Thus, building a set of maps at consecutive instants entails capture, storage and transfer of huge number of images. Data compression becomes a mandatory step for the DSM implementation.

Compression of coherent optical signals can be done for real and complex-valued signals [22-24]. In the intensity-based DSM, different algorithms are applied to intensity data [4, 25-29]. Relevant information is not the intensity value but its change in time. Based on this DSM feature, we proposed and analyzed several approaches for lossy data compression. As a first solution, we transformed 8-bit encoded speckle images into binary images with only two intensity levels [30]. We applied as a binarization threshold the 2D distribution of the pointwise estimate of the averaged in time intensity calculated for each pixel. Using this pointwise threshold, we obtained activity maps, which were as informative as the ground truth map (GTM) computed from 8-bit encoded bmp images. The advantage of binarization is its efficiency in processing for non-uniform illumination when the speckle has spatially varying statistics. The drawback is the need for preprocessing before the binarization. We continued studying speckle data compression by analysis of the coarse quantization of intensity [31, 32]. We studied the case when the data were quantized directly at acquisition without calculation of the average intensity and normalization. We considered as coarse any quantization with fewer levels than at acquisition, e.g. less than 256 levels. Efficiency of the coarse quantization depends on the intensity histogram within the acquired images. We proposed usage of uniform scalar quantization for symmetric intensity distributions when the speckle contrast is low. We proved efficiency of the non-uniform quantization for an asymmetric intensity distribution observed for a high-contrast speckle and uniform illumination. We studied also the case of non-uniform illumination when the histogram of intensity is asymmetric, the average and the variance of the intensity vary across the object and it is necessary to introduce normalization in the processing algorithm after the quantization. In all analyzed cases, we obtained activity maps comparable to the GTM for a number of levels going down from 256 to 8 or 16 depending on the tested objects.

This paper addresses applicability of two common JPEG compression standards, JPEG and JPEG2000, to speckle images acquired in the intensity-based DSM. Both standards are well recognized and supported and provide significant data compression in image processing. The conventional JPEG format may still boast with a dominant use while JPEG2000 is well accepted for medical image coding [33]. Both standards are realized by different algorithms and so exhibit different artifacts. Despite the substantial differences between them, they distort the recorded images in a similar way from the DSM point of view, i.e. they change the spatial correlations within each image. This distortion inevitably affects the activity map. The JPEG Pleno is not considered in the paper as being developed for light field, holography and point cloud data compression [34]. We focus on the lossy compression scheme in view that the original speckle images are not needed for the processing stage and that the lossless compression provides rather modest compression ratios.

The objective of the paper is to prove that the images compressed by using JPEG or JPEG2000 are useful for the pointwise DSM. Analysis of both compression schemes is done by comparing activity maps built from decompressed grayscale synthetic images and color experimental images to the GTMs from images in bmp format. We limit analysis to the case of 256 quantization levels as acceptable for most of the DSM applications. As figures of merit of the studied compression schemes, we use the probability density function (PDF) of the estimate of the statistical parameter characterizing the activity as well as the structural similarity index (SSI) between the maps built from decompressed and original images. We study compression of low and high contrast speckle for uniform/non-uniform illumination.

The paper is organized as follows: in Sec.2 we give brief description of the intensity-based pointwise DSM and analyze distortions in the decompressed speckle images stored in JPEG and JPEG2000 formats. The study presented in Sec.3 focuses on distortions induced in the activity maps. Both Sec.2 and Sec.3 present results for simulated grayscale speckle images. Applicability of both compression schemes is analyzed in Sec.4 for the case of experimentally acquired images in two dynamic speckle measurements.

## 2. DISTORTION ANALYSIS OF DECOMPRESSED SPECKLE IMAGES

### A. Intensity-based pointwise dynamic speckle method

For convenience, we briefly explain the capture and processing steps of the intensity-based pointwise DSM. The capture is shown in Fig.1. A 3D object on a vibration-isolated table is illuminated with linearly polarized laser light. The light beam is spatially filtered, expanded and collimated. A 2D optical sensor is focused on the object and captures the light reflected from the object surface. The processes in the object cause phase changes in the complex amplitude of light, which enters the sensor aperture. The sensor records images of fluctuating in space and time intensity distributions. The images have $N_x \times N_y$ pixels at a pixel interval, $\Delta$. The time interval, $\Delta t$, between two consecutive images provides correlation of intensities in a sequence $I_{kl,n} \equiv I(k\Delta, l\Delta, n\Delta t), n = 1..N$ recorded at pixel $(k\Delta, l\Delta)$ at $N$ instants. The images are stored using JPEG formats.

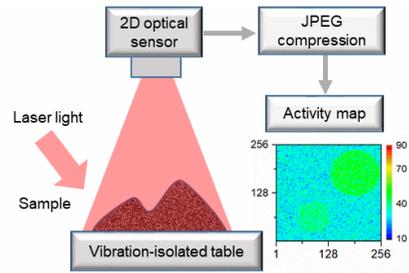

Fig. 1. Schematic representation of intensity-based DSM.

The processing step of the method is given in Fig.2. A single sequence of $N$ correlated speckle images is used to calculate an activity map. A set of $P$ such sequences for instants $t_1, t_2....t_P$ is processed with $t_i$ being the time instant of recording the first image in the $i$-th sequence. Then, $P$ activity maps are produced for the processes ongoing in the object. The choice of $t_1, t_2....t_P$ depends on the task, so the image sequences may overlap or be separated by equal/unequal time intervals. In the 3D volume $N_x \times N_y \times N$ of speckle data in each image sequence, the data are weakly correlated in space and show some spatial distribution of the temporal correlation radius, $\tau_c(k,l) = \tau_c(k\Delta, l\Delta)$, of intensity fluctuations. The larger the temporal radius, the larger the correlation, the slower the process and the lower activity. Different correlation-based algorithms are applied to build the set of activity maps to show evolution of $\tau_c(k,l)$ in time. We give in Fig.2 activity maps for a synthetic object with two circular regions surrounded by a background with much slower intensity fluctuations at $\tau_c$ = 50 $\Delta$t. The maps were calculated with the modified structure function (MSF) [28]:

$$S_1(k,l,m) = \frac{1}{(N-m)} \sum_{i=1}^{N-m} |I_{kl,i} - I_{kl,i+m}| \quad (1)$$

where the integer $m$ shows the time lag, $\tau = m\Delta t$, between the compared intensities. The radius $\tau_c(k,l)$ for the larger circular region takes values of 8, 12 and 20 Δt at $t = t_1$, $t_2$ and $t_P$; for the smaller region, $\tau_c$ is equal to 14, 20 and 30 Δt at these moments. The time lag is τ = 10 Δt. The number of images is 256. They have been simulated as 8-bit encoded bitmap images as is described below for $N_x \times N_y$ = 256×256 and a wavelength of 532 nm. The maps clearly demonstrate the spatial regions of different activity and the MSF decrease at lower activity.

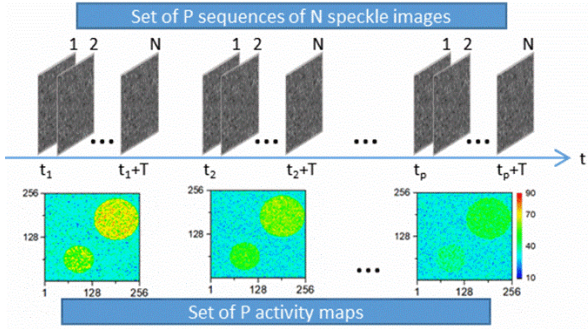

Fig. 2. Capture of a set of sequences of speckle images at instants $t_1$, $t_2$ and $t_P$ (top) and computation of a set of MSF maps related to these instants (bottom); the maps are plotted with the same color scale and show decrease of the MSF from left to right.

**B**. **Distortions in decompressed synthetic speckle patterns**

Effective way to study the impact of various compression schemes is processing of synthetic speckle patterns for controllable spatial distribution of activity. In simulation, we accepted the model of scattering centers, which changed their positions normally to the object surface. Mutual independence was assumed between the amplitude and the phase of light scattered from a given center and between the amplitudes and phases at any two centers. Thus, the normally distributed phase change, $\Delta\varphi_m^{kl}$, at point $(k\Delta, l\Delta)$ between the moments separated by a time lag $\tau = m\Delta t, m = 1,2...N_\tau < N$ leads to temporal fluctuations of intensity in the optical sensor with a normalized correlation function $\rho_{kl}(\tau = m\Delta t) = \exp(-\sigma^2\{\Delta\varphi_m^{kl}\})$ [35]. Here $\sigma^2\{\Delta\varphi_m^{kl}\}$ is the variance of the phase change. Different models can be used for $\rho_{kl}(\tau = m\Delta t)$, but we chose an exponentially decreasing function $\rho_{kl}(\tau) = \exp[-\tau/\tau_c(k,l)]$ as appropriate for description of many processes. The standard deviation of the phase variation between any two successively captured images is given by $\sigma\{\Delta\varphi_{m=1}^{kl}\} = \sqrt{\Delta t/\tau_c(k,l)}$.

The simulation included the following steps:

Step 1: generation of 2D spatial distributions of delta-correlated random phase on the object surface $\varphi(k\delta, l\delta, i\Delta t), k = 1..2N_x, l = 1..2N_y, i = 1..N$ with $\delta = \Delta/2$ by using a 2D array of random phase values with uniform distribution from 0 to $2\pi$.

Step 2: computation of the phase distribution at $i\Delta t$ from
$$\varphi(k\delta, l\delta, i\Delta t) = \varphi[k\delta, l\delta, (i-1)\Delta t] + \sqrt{\Delta t/\tau_c(k,l)} N(0,1) \quad ,$$

where $N(0,1)$ is a newly generated for each value of "$i$" normally distributed random number with zero mean and variance equal to 1, $k = 1..N_x, l = 1..N_y, i = 1..N$.

Step 3: generation of the complex amplitude on the object surface for intensity distribution $I_0(k\delta, l\delta)$ of the laser beam at the instant, $i\Delta t$, $U_S = \sqrt{I_0(k\delta, l\delta)} \exp\{-j\varphi(k\delta, l\delta, i\Delta t)\}$.

Step 4: generation of the complex amplitude of the light field on the sensor aperture $U_{cam} = FT^{-1}\{H \cdot FT\{U_S\}\}$ with $H$ given by a *circ* function for a diffraction limited $4f$ capture system [36] and $FT\{\cdot\}$ corresponding to Fourier transform.

Step 5: summation of intensity values $|U_{cam}|^2$ in a window of size 2×2 pixels for simulating integration of speckle by the camera pixels; time averaging of the speckle at acquisition is not simulated.

Step 6: recording of the generated speckle images of $N_x \times N_y$ pixels at a pixel interval, $\Delta$, as 8-bit encoded grayscale images in bitmap format (file extension bmp), JPEG format (file extension jpg) at a given quality, Q, and JPEG2000 format (file extension jp2) at a given compression ratio, $\eta$.

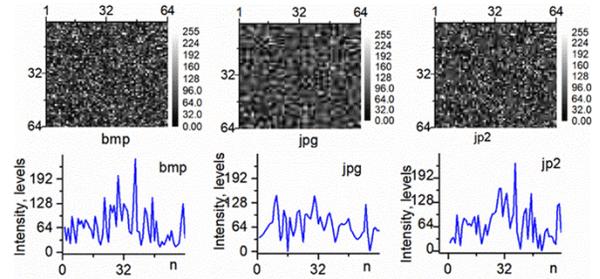

Fig. 3. Sections of synthetic grayscale speckle images (64×64 pixels) in bmp, jpg and jp2 formats (top) and intensity distributions along column 32 in the presented image sections.

Generation of the same sequence of correlated in time speckle images in the three formats allowed for adequate comparison of the activity maps from the compressed data to the GTMs from bitmap images. The histogram of intensity in the original bitmap image is adjusted to cover the entire range of 256 levels. The wavelength was 532 nm. Analysis of compression efficiency was done for symmetric (low contrast speckle) and asymmetric (high contrast speckle) intensity distributions in the speckle images. The speckle contrast was controlled by the transfer function, $H$. We performed step "4" simultaneously for uniform and Gaussian beam, $I_0(k\Delta, l\Delta) = \exp\{-[(k - N_x/2)\Delta^2 + (l - N_y/2)\Delta^2]/\Omega^2\}$, $k = 1,2...N_x, l = 1,2...N_y$. The parameter $\Omega$ gives the spread of the laser beam on the object surface. Study of the case of non-uniform illumination causing spatial variation of speckle statistics is critical for strongly compressed images. For non-uniform illumination, the histogram of the intensity within the images is asymmetric independently of the speckle contrast, and it is not an estimate of the intensity PDF.

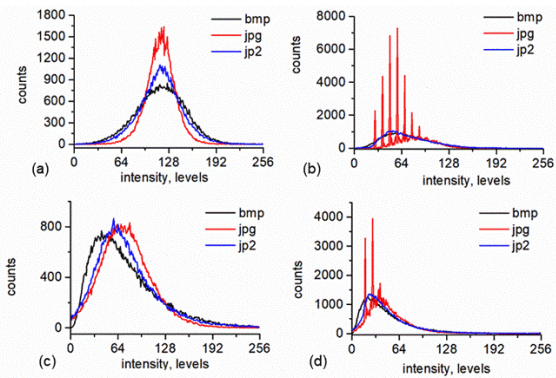

Fig. 4. Histograms of intensity distributions within a grayscale bmp speckle image with 256×256 pixels and its jpg and jp2 versions with 6 times smaller size; (a) low contrast speckle, uniform illumination, (b) low contrast speckle, Gaussian illumination, (c) high contrast speckle, uniform illumination, (d) high contrast speckle, Gaussian illumination.

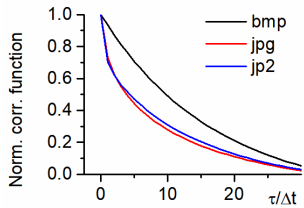

Fig.5. Estimates of the normalized temporal correlation function of intensity fluctuations for a sequence of 256 images recoded in bmp, jpg and jp2 formats; high contrast speckle, uniform illumination.

In order to study distortions introduced by compression, we focused first on distortions in the speckle images. For the purpose, we generated images of size 256×256 pixels and recorded them at different degree of compression. We compared in Fig.3 a section of an 8-bit encoded grayscale bitmap image with asymmetric intensity distribution for uniform illumination to its jpg version recorded with Q = 10 and its jp2 version recorded at $\eta = 6$. Both compressed images are 6 times smaller in size (about 10 KB) than the size of the bitmap image (65 KB). The intensity distributions along column 32 in the presented sections are also given in Fig.3. We especially chose to depict in Fig.3 the case of strongly compressed grayscale image. As is seen, the image decompressed from the jpg format clearly shows the typical for this type of compression blocking artifact. The section decompressed from the jp2 image shows greater resemblance to the bmp section.

We built the intensity histograms for the bmp and the decompressed images. Again, we present in Fig.4 the results only for the case of high degree of compression (Q = 10 and $\eta$ = 6). The histograms are shown for low and high speckle contrast under uniform and Gaussian illumination. For the latter case, we took $\Omega = 400\Delta$ for the image size 256×256 pixels. This choice of $\Omega$ corresponds to rather small drop of the intensity at the image periphery compared to the center. Nevertheless, even such a slight non-uniformity severely increased the blocking artifacts in the jpg images at large degree of compression as is seen from Fig.4(b) and Fig.4(d).

The JPEG and JPEG2000 compression schemes change the temporal correlation within the intensity sequences used to form the entries for an activity map. We generated 256 speckle images with the same correlation radius in all points, $\tau_c = 20\Delta t$, and estimated the mean normalized temporal correlation function from

$$\hat{\rho}(m) = \frac{1}{N_x N_y (N-m)} \sum_{k=1}^{N_x} \sum_{l=1}^{N_y} \frac{1}{\sigma_{kl}^2} \sum_{i=1}^{N-m} (I_{kl,i} - \bar{I}_{kl})(I_{kl,i+m} - \bar{I}_{kl}) \quad (2)$$

$$\sigma_{kl}^2 = \frac{1}{N} \sum_{i=1}^{N} (I_{kl,i} - \bar{I}_{kl})^2 \ , \quad \bar{I}_{kl} = \frac{1}{N} \sum_{i=1}^{N} I_{kl,i} \quad (3)$$

The images were generated both for speckle with low and high contrast, under uniform and Gaussian illumination. Note that the estimate (2) is biased with respect to the real normalized function, $\rho_{kl}(\tau) = \exp[-\tau/\tau_c(k,l)]$, which is the same for all points. The bias is due to determination of the estimates (3) of the mean value, $\bar{I}_{kl}$, and the variance, $\sigma_{kl}^2$, at each point from a finite and rather short with respect to $\tau_c$ sequence of intensity values. We give in Fig.5 falling of $\hat{\rho}(m)$ with the time lag for bmp images, jpg images with Q = 10 and jp2 images with $\eta$ = 6. The presented curves correspond to the high contrast speckle under uniform illumination. The curves for the low contrast speckle are practically the same. The same is true for Gaussian illumination taking into account that normalization in (2) is pointwise. The result in Fig.5 proves that both JPEG schemes change the temporal correlation for the points of successively acquired correlated speckle images.

## 3. DISTORTIONS IN ACTIVITY MAPS FROM DECOMPRESSED SPECKLE IMAGES

### A. Analysis at constant activity

Raw speckle data lead to strong fluctuations of the activity estimates from point to point and spread their PDFs, thus worsening sensitivity of the method. If simulation is made at spatially constant activity, which means to have the same value of the temporal correlation radius in all pixels, $\tau_c(k,l) = const$, it is possible to build a histogram of the estimate used for activity evaluation. For an image containing 256×256 points, the histogram is built from 65536 entries and can be considered as a good approximation to the PDF of the estimate. We simulated sequences of correlated images at constant activity to study the impact of data compression on the PDFs of the estimates. In Fig.6, we plotted the histograms obtained from activity maps built at a time lag $\tau = 10 \Delta t$ from decompressed jpg images at Q = 70 (image size 32.4 KB), 30 (19.5 KB) and 10 (9.77 KB) and from decompressed jp2 images at $\eta$ = 2 (31.4 KB), 3 (20.7 KB) and 6 (10 KB). In this way, we compare the maps and the histograms for compressed images of almost equal size. The size of the bmp images was 65 KB. For all simulated cases, $\tau_c = 20 \Delta t$. The presented histograms correspond to processing of high contrast speckle for illumination with 532 nm. The histograms in Fig.6 (a) and Fig. 6 (b) correspond to uniform illumination and hence to the estimate $S_1$ given by Eq. (1).

This algorithm, however, is not appropriate for the case of Gaussian illumination and we used the following algorithm:

$$S_2(i,k,m) = \frac{1}{(N-m)} \sum_{i}^{N-m} \frac{|I_{kl,i} - I_{kl,i+m}|}{(I_{kl,i} + I_{kl,i+m} + q)} \quad (4)$$

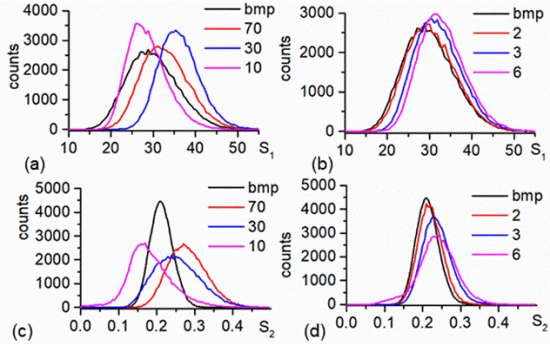

Fig. 6. Histograms of the activity estimates for uniform illumination, $S_1$ (a,b), and Gaussian illumination, $S_2$ (c,d), determined from bmp images and decompressed JPEG (a,c) and JPEG2000 (b,d) images at different degrees of compression and constant activity with $\tau_c = 20\Delta t$; $N = 256$, $\tau = 10\Delta t$.

The parameter $q$ in (4) stabilizes the algorithm for strongly compressed data. We used $q = 1$ in the results presented below. The parameter $\Omega$ was equal to $400\Delta$. As is seen, both compressing schemes shift the histograms of the activity estimates with respect to the histogram obtained for the bmp images. As a whole, compression narrows the histograms for the estimate $S_1$ and vice versa for $S_2$. The compression impact is greater for the JPEG compression scheme.

B. **Analysis at spatially varying activity**

The objective of the DSM is to indicate areas of faster or slower changes on the object surface due to some processes of various origin. The studied compression schemes modify substantially the recorded 8-bit images. Given the fact that absolute values of intensity are not important, the lossy data compression is acceptable if it keeps information about i) spatial variation of the speed of the processes across the object and ii) evolution of this speed in time. We first analyzed efficiency of lossy JPEG or JPEG2000 compression for the task of visualization of spatial distribution of activity. This was done by simulation for an object with two compact regions of a rapidly evolving process that are buried in a background with slower temporal variation of intensity. As in our previous studies [32], we formed the higher activity regions as the logos "IOMT" and "ETRI" of both research institutions involved in the current analysis. A test object thus created is suitable for checking efficiency of the DSM for detection of relatively small areas of higher activity with sharp borders. We simulated both high and low contrast speckle, but here we present only the high contrast case. The simulation parameters were as follows: temporal correlation radii for the logos and the background equal to $\tau_{cl} = 10\Delta t$ and $\tau_{cb} = 40\Delta t$ respectively, time lag $\tau = 10\Delta t$ ($m = 10$), image size 256×256 pixels, $N = 256$, wavelength 532 nm. The maps from the bmp images are the GTMs. They are shown in Fig.7 for uniform and Gaussian illumination and the estimators $S_1$ and $S_2$. For illustration, we included in Fig.7 (b) the map obtained with $S_1$ from patterns acquired under non-uniform illumination. No smoothing was applied to the maps to decrease the fluctuations. We see that the maps in Fig.7 (a) and Fig.7 (c) correctly reflect activity in the object. The sharp borders of the logos regions are also well reconstructed due to pointwise processing. The result in Fig.7(b) reflects the non-uniform intensity distribution in the laser spot.

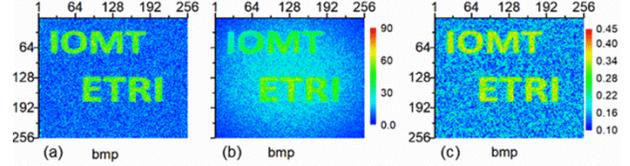

Fig. 7. Ground truth activity maps for an object with two logos and uniform background: (a) $S_1$, uniform illumination, (b) $S_1$, Gaussian illumination, (c) $S_2$, Gaussian illumination, $\tau_{cl} = 10\Delta t$, $\tau_{cb} = 40\Delta t$, $N = 256$, $\tau = 10$ $\Delta t$.

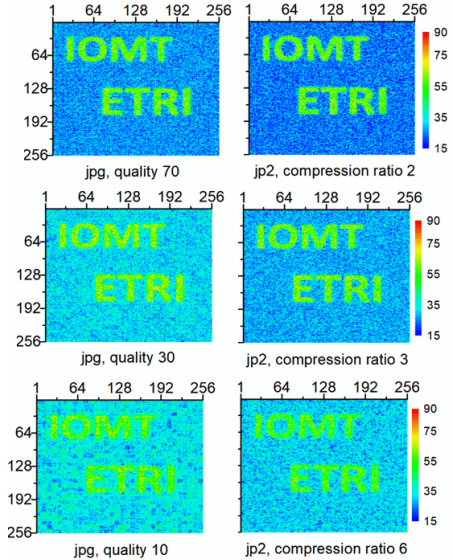

Fig. 8.
Activity maps from decompressed images for an object with two logos and a uniform background under uniform illumination; type and degree of compression are given under each map, $\tau_{cl} = 10\Delta t$, $\tau_{cb} = 40\Delta t$. $N = 256$, $\tau = 10$ $\Delta t$.

The maps from the decompressed images for uniform illumination are shown in Fig.8 for 2, 3 and 6 times smaller size of the compressed images compared to the bmp images. For better evaluation of the compression impact, we computed maps of the SSI (Fig.9) between the activity maps from bmp and decompressed jpg and jp2 images. The same scale of the maps in Fig.8 and Fig.9 helps to visualize better the impact of degree of compression. The higher activity in the regions of the logos is properly detected even at large degree of compression. Decreasing the image size twice by applying JPEG or JPEG2000 keeps rather high the similarity to the GTM for the activity maps from the decompressed images. Further decrease of the image size leads to artifacts in the activity maps, especially to blocking artifact typical for JPEG compression. The similarity of the activity map to the GTM is better for JPEG2000 algorithm and is greater in the regions of higher activity (regions of the logos). It remains comparatively high in these regions even for low quality of JPEG compression or high compression ratio of JPEG2000. The mean SSI for comparing the maps from decompressed jpg images with Q = 70, 30 and 10 to the GTM is equal to 0.955, 0.783 and 0.503, respectively. The same index in the case of jp2 format with η = 2, 3 and 6 is equal to 0.992, 0.925 and 0.639, respectively. Actually, similarity between the GTM and the map from decompressed images is not so critical. The vital requirement is to differentiate between the regions of different activity. The observed artifacts are stronger in the background region but they do not interfere with correct description of activity for the used test object. This means that lossy JPEG or JPEG2000 compression is fully appropriate for tasks dedicated to visualization of spatial regions with different speed of the ongoing processes.

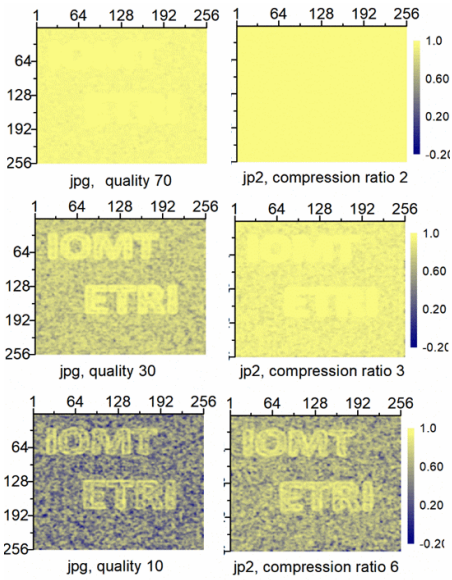

Fig. 9. Maps of SSI for comparison of activity maps from decompressed and bmp images for an object with two logos and uniform background under uniform illumination; type and degree of compression are given under each map, $\tau_{cl} = 10\Delta t$, $\tau_{cb} = 40\Delta t$, $N = 256$, $\tau = 10\ \Delta t$.

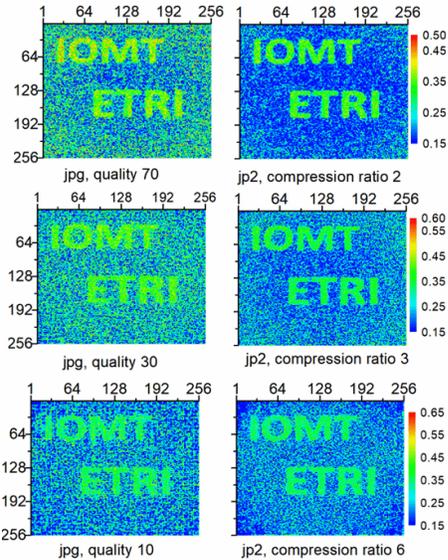

Fig. 10. Activity maps from decompressed images for an object with two logos and a uniform background under Gaussian illumination; type and degree of compression are given under each map, $\tau_{cl} = 10\Delta t$, $\tau_{cb} = 40\Delta t$, $N = 256$, $\tau = 10\ \Delta t$.

The results of normalized processing of decompressed images for Gaussian illumination are shown in Fig.10. The reconstruction of the logos is good, even when the image size decreases from 65 to 10 KB. The blocking artifacts are removed by normalization in Eq. (4). The mean SSI for comparison of the maps computed from jpg images with Q = 70, 30 and 10 to the GTM is 0.645, 0.467 and 0.410, respectively. Compared to uniform illumination, the SSI decrease is greater. This index is again larger for the JPEG2000 compression: it achieves 0.881, 0.646 and 0.486 for η = 2, 3 and 6, respectively. Despite increasing dissimilarity of the decompressed jpg and jp2 images to the original bmp images for non-uniform illumination, lossy compression using jpg and jp2 formats is completely applicable for the DSM.

## 4. COMPRESSION OF EXPERIMENTAL SPECKLE IMAGES

We proved applicability of JPEG or JPEG2000 compression in real dynamic speckle measurements by processing data from two drying experiments. The first experiment was conducted with a metal coin covered by a non-transparent paint. The main advantage of this object is the complicated relief formed by grooves and embossments. Different speed of paint evaporation for these relief's elements and the flat coin surface makes possible effective evaluation of the detrimental effect of compression on activity visualization. The second experiment included capture of speckle images for drying of a droplet from water and methanol solutions of the azopolymer poly [1- [4- (3-carboxy-4-hydroxy-phenylazo) benzene-sulfonamido]-1,2-ethanediyl, sodium salt] at different temperatures. We use abbreviation PAZO as a short name for this polymer from Sigma Aldrich. We use thin layers of this anisotropic material for polarization holographic recording. The DSM is highly suitable for monitoring of drying of deposited thin layers of PAZO. The observed droplet was an object with a smoothly changing thickness. The activity maps obtained for the droplet may be spatially inhomogeneous and may evolve in time. This makes the drying droplet a suitable object for checking whether processing of the decompressed jpg or jp2 images provides the same results as the obtained from the bmp images.

The experiments were done using the acquisition set-up in Fig.1. Color CMOS camera X06c-s (Baumer) recorded images with 780×582 pixels at pixel pitch 8.3 μm. The images were recorded with exposure time 20 μsec at Δt equal to 250 msec. The environmental temperature was 25°C. A He-Ne laser emitting at 632.8 nm illuminated the objects on a vibration-isolated table. Linear polarization of the light was checked with PAX5710VIS-T polarimeter (Thorlabs).

We present here the results obtained for the PAZO water solution. For the experiment, 20 mg of PAZO were dissolved in 400 μl water to obtain the concentration used usually for deposition of PAZO thin films. A 10 μl droplet was spread on a microscope glass slide, placed on a hot stage THMS 600 (Linkam Scientific). The stage kept the object temperature at a pre-set value. The glass slide stayed for 5 minutes on the hot stage to reach thermal equilibrium before deposition of the droplet. We recorded 10 sets of 256 speckle patterns each at 30, 40, 50 and 60°C with 2 minutes interval between two consecutively recorded sets. The measurement at each temperature was done with a new droplet. Exemplary speckle images for the coin and the droplet of a polymer solution in the bmp format and after decompression of the jpg and ip2 versions are given in Fig.11. The images are plotted with the same scale. The size of the bmp image is 1.29 MB, the jpg and ip2 images are 138 KB and 132 KB respectively, i.e. about 10 times smaller.

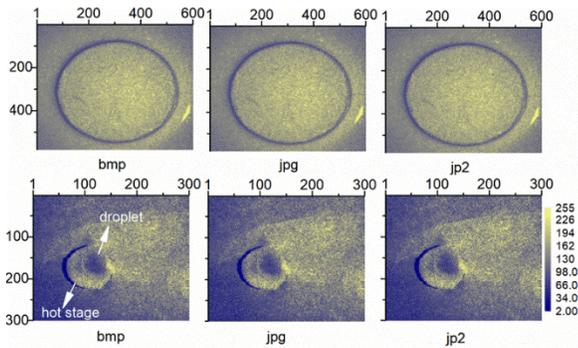

Fig. 11. Image plots for speckle data recorded in bmp, jpg and jp2 formats in the drying experiments with a coin covered with paint (top) and a droplet of a polymer solution on a hot stage (bottom); wavelength 632.8 nm.

The GTM and the maps from the decompressed jpg and jp2 images in the coin's experiment are shown in Fig.12 and Fig.13. As is seen in Fig.11, the coin surface in not uniformly illuminated, and to extract properly the activity map, we used the normalized estimate

$$S'_1(k,l,m) = \frac{1}{(N-m)} \sum_{i=1}^{N-m} \frac{1}{\sigma_{kl}} |I_{kl,i} - I_{kl,i+m}| \qquad (5)$$

where the estimate of the standard deviation, $\sigma_{kl}$, is given by formula (3). This normalization provided better results than the estimate (2). The files in jpg format were recorded at Q = 95, 90, 70, 50 and 30 thus producing compressed image sizes equal to 138 KB, 80.8 KB, 36.4 KB, 23.9 KB and 15.0 KB respectively. The obtained activity maps resemble with good spatial resolution the coin relief. Note that this is an activity map, and not a 3D reconstruction of the object. The activity maps are satisfactory for Q ≥ 50. Compression is about 55 times at Q = 50. Further decrease of the compressed image size worsens the map beyond acceptance. The chosen normalization of the estimate masks the blocking artifacts.

The activity maps for jp2 in Fig.13 correspond to η = 20, 30, 40, 50 and 60. The maps are plotted with the same scale as in Fig.12: the normalized estimate (5) varies from 0.7 to 1.3. The maps at the highest compression ratios, η = 50 and 60, exhibit acceptable quality of visualization. As a whole, one may conclude that the JPEG2000 outperforms the JPEG for activity visualization. Both compression schemes provide good activity maps at rather high degree of compression

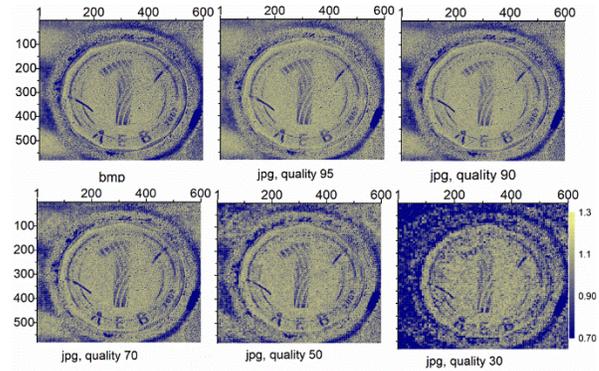

Fig. 12. Activity maps for a coin covered with paint; maps are built from 256 decompressed jpg images recorded with different quality and compared to the GTM from images in the bmp format; τ = 10Δt.

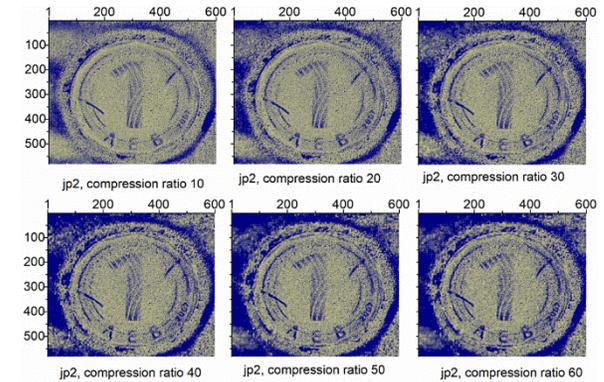

Fig.13. Activity maps for a coin covered with paint; maps are built from 256 decompressed jp2 images recorded with different compression ratios and compared to the GTM from images in the bmp format; τ = 10 Δt

We computed the activity maps for the droplet for all sets of images acquired at 30, 40, 50 and 60°C using the estimates $S_1$ and $S_2$. The second estimate is more appropriate in the case due to non-uniformity of intensity in the laser spot and lower reflectivity in the droplet. However, we present in the paper the non-normalized estimator to visualize the artifacts induced by compression. In Fig.14, we compare the GTM for one of the sets acquired at 40°C with the maps from the decompressed jpg and jp2 images with a 10, 20 and 30 times smaller size. The maps from the decompressed images are informative independently of the observed artifacts. The mean SSI for comparing the maps from the decompressed images and the GTM is 0.924, 0.846 and 0.739 for the jpg compression and 0.941, 0.892 and 0.791 for jp2 images at 10, 20 and 30 times decrease of the size.

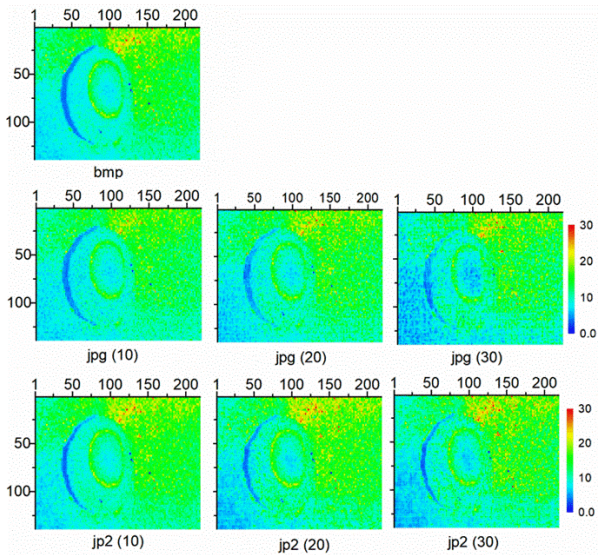

Fig. 14. Comparison of $S_1$ activity maps from decompressed images and from bmp images for a droplet of polymer solution; the numbers in the brackets show the ratio between the bmp and the compressed image sizes.

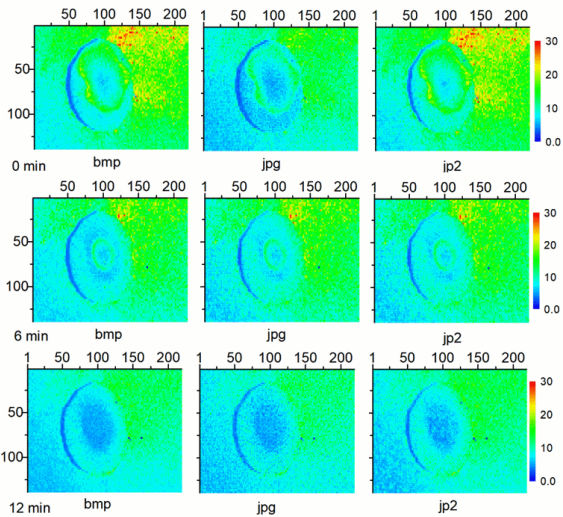

Fig. 15. Comparison of $S_1$ activity maps from decompressed images and from bmp images for a droplet of polymer solution at compression ratio equal to 10; the time in minutes show the start of acquisition with respect to the beginning of the experiment.

For all sets of speckle images acquired at 30, 40, 50 and 60°C, we determined the time dependence of the average MSF estimate $S_1$ at lag $\tau = 10\ \Delta t$ by averaging in the area of 100 by 100 pixels on the activity maps around the center of the droplet. This was done for the maps from sequences of 256 bmp and decompressed images of 10 and 20 times smaller size. The 10 000 values of the estimate in the chosen spatial area lead to smooth decrease of the average estimate with time. For illustration of activity maps evolution, we present in Fig.15 the maps from bmp, jpg and jp2 images at three starting times, $t_1$, $t_2$ and $t_3$, of image acquisition. The images were acquired at 60°C. In Fig.15, the size of the compressed images is 10 times smaller than the size of the bmp images equal to 1.29 MB. At $\eta = 10$, the maps from the decompressed images strongly resemble the GTM. We plotted in Fig.16 the average MSF estimate as a function of time for 60°C. The values of the average MSF for the bmp and jpg images are rather close, although the difference between the curves increases at $\eta = 10$ and decrease at $\eta = 20$. For the jp2 compressed images, the averaged MSF is higher than for the bmp images. The difference between the curves for the bmp and jp2 images increases with $\eta$ and slightly with time.

Despite the difference between the bmp and jp2 curves, this compression scheme enables more correct characterization of the polymer droplet drying process, because the relative changes of the estimate and not its absolute value are informative.

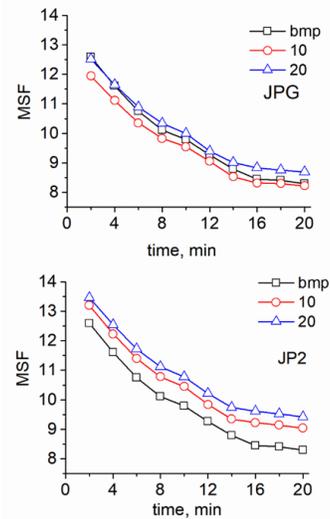

Fig. 16. Drop of activity estimate $S_1$ in time for a drying droplet of a polymer solution at 60°C for processing of bmp and decompressed jpg (top) and jp2 (bottom) images at compression ratios 10 and 20.

## 5. CONCLUSIONS

In summary, we proposed JPEG and JPEG2000 lossy compression in 2D intensity-based dynamic speckle analysis. The input data for this analysis are sequences of correlated in time speckle images acquired for laser illuminated objects. The output is a 2D activity map, which shows the regions on the object surface of fast or slow intensity changes or the speed of a process causing these changes, respectively. The sequence needed for a single activity map comprises dozens of speckle images. A large number of maps is required to analyze evolution of a process. Data compression becomes more than necessary especially taking in view data redundancy characterizing the DSM. How effective JPEG and JPEG2000 formats are for compressing dynamic speckle data is not a trivial question. On the one hand, this analysis is predominantly qualitative: it simply indicates areas with different speed of processes. As we have proven by processing synthetic and experimental speckle data, both

JPEG formats provide high quality of activity visualization at comparatively high compression ratios, e.g. 50 times decrease of size for color images. On the other hand, the DSM has a potential for quantitative characterization. Being transform-based approaches, JPEG and JPEG2000 change the temporal correlation between the intensity values at a point to the extent, which depends on the compression ratio. This inevitably changes the functional dependence of the estimate statistics on time when observing a certain process. We limited our analysis of this issue to evaluating the change inflicted by compression on the average value of the activity estimate for a drying droplet of a polymer solution. We observed change in the time-dependence of the average with respect to the result for the bmp images, but this change was rather small. Given the significant gain in computer memory provided by the JPEG or JPEG2000 compression, our assessment is that distortion of time dependencies is of secondary importance. We plan, however, to study more thoroughly the impact of JPEG compression on statistical properties of the estimates for the case of varying in time activity. The main result of the performed study is that JPEG and JPEG2000 formats can be recommended for storage of images in the DSM. Similarly to medical imaging, JPEG2000 slightly outperforms the conventional JPEG.

**Funding** Institute of Information and Communications Technology Planning and Evaluation (IITP) grant funded by the Korea Government (MSIT) (2019-0-00001, Development of Holo-TV Core Technologies for Hologram Media Services)

**Acknowledgments** E. Stoykova thanks European Regional Development Fund within the Operational Programme "Science and Education for Smart Growth 2014–2020" under the Project CoE "National centre of Mechatronics and Clean Technologies" BG05M2OP001-1.001-0008. M. Levchenko thanks 2020 Plenoptic Imaging project for supporting his PhD training. This project has received funding from the European Union's Horizon 2020 research and innovation programme under the Marie Skłodowska-Curie grant agreement No 956770.

**Discloueres** The authors declare no conflicts of interest related to this article.

**Data Availability** Data underlying the results presented in this paper are not publiy available at this time but may be obtained from the authors upon reasonable request.